\documentclass[12pt]{article}

\def\d{\partial}
\def\tr{{\rm tr}}

\begin{document}

\title{Mathematical properties of cosmological models with accelerated 
expansion}

\author{Alan D. Rendall\\
Max-Planck-Institut f\"ur Gravitationsphysik\\ Am M\"uhlenberg 1\\
D-14476 Golm, Germany}

\date{}

\maketitle

\begin{abstract}
An introduction to solutions of the Einstein equations defining cosmological 
models with accelerated expansion is given. Connections between 
mathematical and physical issues are explored. Theorems which have been
proved for solutions with positive cosmological constant or nonlinear
scalar fields are reviewed. Some remarks are made on more exotic models
such as the Chaplygin gas, tachyons and $k$-essence.
\end{abstract}

\section{Introduction}
Recent cosmological observations indicate that the expansion of the 
universe is accelerating and this has led to a great deal of 
theoretical activity. Models of accelerated cosmological expansion
also raise a variety of interesting mathematical questions. The
purpose of the following is to first give a pedagogical introduction
to this subject suitable for the mathematically inclined reader and then
to present an overview of some of the mathematical results which have been 
obtained up to now and the many challenges which remain.

The simplest class of cosmological models consists of those with the
highest symmetry, i.e. those which are homogeneous and isotropic. The
underlying spacetimes are the FLRW (Friedmann-Lemaitre-Robertson-Walker)
models. A further simplification can be achieved by assuming that the 
metric of the slices of constant time is flat. The spacetime metric can be 
written in the form:
\begin{equation}
-dt^2+a^2(t)(dx^2+dy^2+dz^2)
\end{equation}      
for a suitable scale factor $a(t)$. These are the models most frequently 
used in the literature due both to their simplicity and the fact that 
spatially flat FLRW models appear to give a good description of our 
universe.

The physical interpretation of $a(t)$ is that if two typical galaxies
are a distance $D(t)$ apart at time $t$ then $D(t_2)/D(t_1)=a(t_2)/a(t_1)$
for any times $t_1$ and $t_2$. The statement that the universe is
expanding corresponds to the condition that the time derivative $\dot a$ is
positive. Accelerated expansion means that the second derivative $\ddot a$ is
positive.

The function $a(t)$ should be determined by the field equations for  
gravity and in the following we always take the Einstein
equations for this purpose. There are two choices to be made. The one
concerns the cosmological constant $\Lambda$. The other concerns the 
description of the matter content of spacetime. This means
choosing the variables which describe the matter, the equations of 
motion these are to satisfy and the definition of the energy-momentum
tensor as a function of the matter fields and the spacetime metric.
Under the assumption of FLRW symmetry this will lead to an evolution 
equation for $a(t)$. The easiest way to produce models with accelerated 
expansion is to choose a positive cosmological constant ($\Lambda>0$). 
A more sophisticated alternative is to choose $\Lambda=0$ but to
include a suitable nonlinear scalar field among the matter fields. 

The rest of this article is structured as follows. It starts with a
brief introduction to some physical ideas relevant to accelerated
cosmological expansion. Then mathematical theorems about spacetimes
with positive cosmological constant motivated by the physics are described. 
After that these results are compared with the original physical motivation.
Once the case of a positive cosmological constant has been described it
is discussed why it might be good to replace the cosmological constant
by a nonlinear scalar field and what changes when that is done. Finally,
some future research directions involving more general models for
cosmic acceleration are indicated. In particular, comments are
made on the Chaplygin gas and the tachyon condensate.

\section{Physical background} 

Accelerated expansion plays a role in cosmology in two different regimes.
The first is the very early universe while the second is the period 
between the decoupling of the microwave background radiation and the present. 
Accelerated expansion in the early universe is associated with the name
inflation which was introduced by Guth \cite{guth}. The paper \cite{guth}
was extremely important in popularizing the concept of inflation. Related
ideas had been considered previously by other authors. 

One of the attractive features of inflation is that it is claimed to
solve certain \lq problems\rq\ in cosmology. It is justified to ask in which
sense these can really be considered as problems but these philosophical
questions will not be entered into in the following. Among these issues 
are 
\begin{itemize}
\item
homogeneity and isotropy
\item
flatness problem
\item
horizon problem
\end{itemize}
The first issue is that, after averaging on a suitable scale, our universe 
is homogeneous and isotropic. There are two basically different kinds of
possible reason for this. One is that it was always homogeneous and 
isotropic. This possibility is perceived by many as unsatisfactory. The
alternative is that the universe was originally anisotropic and 
inhomogeneous and that some dynamical mechanism later made it homogeneous 
and isotropic. In the second explanation this mechanism must be found.
The second issue is that it appears that the curvature of space on
cosmological scales is very small today and was even smaller at 
decoupling. It is often perceived that this smallness requires an 
explanation. The third issue is that the temperature of the microwave 
background is essentially the same at points such that there would have 
been no time to send a signal to both from some common point since the 
big bang in a standard Friedmann model (without accelerated expansion).
Can this be explained? Inflation has something to say about all three
issues, as will be shown later.

Accelerated cosmological expansion at the present epoque is a relatively
recent discovery, dating from the late 1990's. There is now very strong
observational evidence, which continues to accumulate, that the
velocity of recession of distant galaxies is accelerating. On the
theoretical side this phenomenon is associated with the names dark energy
and quintessence. The latter term was introduced by Caldwell, Dave and
Steinhardt \cite{caldwell}. There are a number of different lines of
evidence for cosmic acceleration at times after decoupling which include
\begin{itemize}
\item
supernovae of type Ia
\item
microwave background fluctuations
\item
gravitational lensing
\item
galaxy clustering
\end{itemize}

Here only the supernova data will be discussed. A supernova of type Ia
is an exploding star which is bright enough to be visible at 
cosmological distances. The characteristics of an event of this type 
which can in principle be observed are the red shift, the light curve 
(observed brightness as a function of time) and the spectrum. In
recent years it has become possible to observe these data in practice
for a useful sample of objects. The light curve and spectrum provide
the information needed to identify a supernova as being of type Ia.
The advantage of this is that type Ia supernovae have universal
properties which allow their intrinsic brightness to be determined.
In a first approximation, all of these objects have the same intrinsic 
brightness at the maximum of their light curves. The number of objects
of this type observed so far is just over 150. The projected space
mission SNAP (Supernova Acceleration Probe) is planned to observe about
2000 per year. The way in which data can be compared with theoretical 
models will be outlined in Section \ref{fluid}.

\section{Mathematical developments}

It has been known for a long time that spacetimes with a positive 
cosmological constant have a tendency to isotropize at late times,
a circumstance associated with the name \lq cosmic no hair theorem\rq.
In \cite{starobinsky83} Starobinsky wrote down formal 
expansions for the late-time behaviour of spacetimes with positive 
cosmological constant. He studied the case where the matter 
is described by a perfect fluid with linear equation of state
$p=(\gamma-1)\rho$, where $\gamma$ is a constant belonging to the
interval $[1,2)$. He also discussed the vacuum case which, as it
turns out, gives the leading order terms in the expansion of the
geometry for the case with fluid as well. In a certain sense the
solutions all look like the de Sitter solution at late times.
This will be made more precise below.

It will be convenient in the following to write the de Sitter
solution in the form
\begin{equation}
-dt^2+e^{2Ht}(dx^2+dy^2+dz^2)
\end{equation}
where $H=\sqrt{\Lambda/3}$. These coordinates only cover half of de
Sitter space but this is no disadvantage in the following where the
subject of interest is the limit $t\to\infty$. The expansions of
\cite{starobinsky83} are expressed in terms of Gauss coordinates. In
other words $g_{00}=-1$ and $g_{0a}=0$ where Latin indices are spatial
indices. In the vacuum case the expansion of the spatial metric is
\begin{equation}
g_{ab}(t,x)=e^{2Ht}(g^0_{ab}(x)+g^2_{ab}(x)e^{-2Ht}+g^3_{ab}(x)e^{-3Ht}
+\ldots)\end{equation} 
The fact that the coefficient of $e^{Ht}$ vanishes is a result of
the analysis. Putting $g^0_{ab}=\delta_{ab}$ and setting all the 
other coefficients to zero gives the de Sitter solution.

In \cite{starobinsky83} it is not specified how this infinite series
is to be interpreted mathematically but it is natural to interpret
it as a formal series. This means that there is no assertion that
the series converges or even that it is asymptotic. Recall that
a series as above is called asymptotic if for each positive integer
$M$ there is a positive constant $C_M$ such that 
\begin{equation}
\left|g_{ab}-\sum_0^M g^m_{ab}e^{(2-m)Ht}\right|\le C_Me^{(1-M)Ht}
\end{equation}
In other words, the sum of any finite truncation of the series differs
from the quantity to which it is asymptotic by a remainder of order
equal to the next term beyond the truncation. A convergent series is
asymptotic but not necessarily conversely. At this point in the discussion
it is not even claimed that the above series is asymptotic. It is just a
formal expression which solves the Einstein equations in the sense that
if we substitute it into the Einstein equations and manipulate the infinite 
series according to rather obvious rules all terms cancel.

In \cite{rendall03a} a theorem was proved concerning the above formal
series. To formulate it, let $A_{ab}$ be a three-dimensional Riemannian
metric and $B_{ab}$ a symmetric tensor which is transverse traceless
with respect to $A_{ab}$. This means that $A^{ab}B_{ab}=0$ and
$\nabla^a B_{ab}=0$ where the covariant derivative is that associated
to the metric $A_{ab}$. Given $A_{ab}$ and $B_{ab}$ of this form which
are smooth ($C^\infty$) there exists a unique series of the above form
satisfying the vacuum Einstein equations with $\Lambda>0$ with smooth 
coefficients $g^m_{ab}$ which satisfies the conditions $g^0_{ab}=A_{ab}$ 
and $g^3_{ab}=B_{ab}$. Notice that on the basis of function counting 
these solutions are as general as the general solution of the vacuum 
Einstein equations. For the general solution can be specified by giving 
the induced metric and the second fundamental form on a spacelike
hypersurface and these must satisfy one scalar and one vector equation.
Thus in both cases we have the same type of data and the same number
of constraints which they have to satisfy. In the present case the
constraints are simpler than in the ordinary Cauchy problem. In
\cite{rendall03a} a corresponding theorem was also proved for the case
of the Einstein equations coupled to a perfect fluid with a linear
equation of state. The most difficult part of the proof is to show 
that the Einstein constraint equations are satisfied as a consequence
of the \lq constraints at infinity\rq, i.e. the tranverse traceless
nature of $B_{ab}$ with respect to $A_{ab}$. 

It is desirable to extend the above results about formal power series
and function counting to show that there exists a large class of 
solutions which have asymptotic expansions of the above form and
that these are general in the sense that they include all solutions
arising from a non-empty open set of initial data on a Cauchy surface.
One place to look for such an open set is as an open neighbourhood
of standard data for the de Sitter solution on a hypersurface $t=$const.
In the vacuum case a result of this kind was proved in \cite{rendall03a}
using results of Friedrich \cite{friedrich86}, \cite{friedrich91} on the 
stability of de Sitter space. The corresponding result with a perfect fluid, 
which is what would be desirable for cosmology, is still open. The proofs
in the vacuum case use the conformal field equations. The results can be 
extended in some cases to conformally invariant matter fields but for
other matter fields, including most fluids, it is not at all clear that
the method could work. If the metric is conformally rescaled as in the
conformal method either the rescaled metric or the conformal factor will
for most fluids be non-smooth, involving non-integral powers of the time
coordinate.

Another possible direction in which the existing results could be
extended is to other spacetime dimensions. In the context of formal 
power series of the vacuum Einstein equations with $\Lambda>0$ this
has been done in \cite{rendall03a}. The result is the series:
\begin{equation}
g_{ab}=e^{2Ht}(g^0_{ab}+\sum_{m=1}^\infty\sum_{l=0}^{L_m} 
(g_{ab})_{m-2,l}t^le^{-mHt})
\end{equation}
where $H=\sqrt{2\Lambda/n(n-1)}$ in spatial dimension $n$. For each 
$m$ the quantity $L_m$ is a finite integer. The terms with $l>0$ will
be refererred to as \lq logarithmic terms\rq\ since $t$ is logarithmic
in the expansion parameter $e^{Ht}$. Again it is possible to prescribe
two quantities $A_{ab}$ and $B_{ab}$ which this time have to satisfy an
inhomogeneous version of the transverse traceless condition in general.
The inhomogeneity is determined by $A_{ab}$. The prescribed coefficients
are $g^0_{ab}=A_{ab}$ and $(g_{ab})_{n-2,0}=B_{ab}$. In general
logarithmic terms are required to get a consistent formal expansion.
They can only be avoided if $n$ is odd, $n=2$ or $A_{ab}$ satisfies some
strong restrictions.

At the present time the results on formal asymptotic expansions for higher 
dimensional vacuum spacetimes have not been extended to existence theorems
for all solutions corresponding to a non-empty open set of initial data
on a regular Cauchy surface. It has, however, been proved that there exists
a very large class of solutions of the Einstein equations with asymptotic
expansions as above. Tensors $A_{ab}$ and $B_{ab}$ satisfying the 
constraints at infinity can be prescribed arbitrarily under the
assumption that they are analytic ($C^\omega$). This was proved in
\cite{rendall03a} using Fuchsian techniques. The generality of the solutions
is judged using function counting. These results can probably be extended 
to fluids with linear equation of state in $3+1$ dimensions but this has not 
been worked out.

The above results require no symmetry assumptions. Under the assumption of
spatial homogeneity much more is known. A theorem of Wald \cite{wald}
shows that for spacetimes of Bianchi types I-VIII with positive
cosmological constant and matter satisfying the dominant and strong
energy conditions solutions which exist globally in the future have
certain asymptotic properties as $t\to\infty$.  This implies that
the asymptotics of these spacetimes have some of the properties 
which follow from the asymptotic expansions discussed above. To go further
the matter model must be specified. For matter described by the Vlasov
equation global existence and more refined asymptotics have been proved
by Lee \cite{lee03}. When the matter model is a perfect fluid with linear
equation of state similar results have been proved in \cite{rendall04a}.
These results confirm many of the features expected from the formal
asymptotic expansions. There is also a class of highly symmetric 
inhomogeneous spacetimes with $\Lambda>0$ for which global existence 
and asymptotic properties has been proved for large initial data.
These are solutions of the Einstein-Vlasov system with plane or
hyperbolic symmetry \cite{tchapnda03a}, \cite{tchapnda03b}. 

\section{Mathematics and physics compared}

In all the classes of spacetimes with a positive cosmological constant
which expand forever the available mathematical results all indicate
isotropization at late times. To see the reason for this, introduce
the second fundamental form of the hypersurfaces $t=$const., which in
Gauss coordinates is given by $k_{ab}=-(1/2)\d_t g_{ab}$. It turns out that
the tracefree part of $k_{ab}$ becomes negligible in comparison with 
its trace $\tr k$, which is the mean curvature. Equivalently each eigenvalue 
of the second fundamental form divided by the mean curvature tends to $1/3$
as $t\to\infty$. In the FLRW models these values are exactly equal to $1/3$.
In the terminology more common in general relativity the ratio of shear
to expansion tends to zero. This is the meaning of isotropization.

At first sight it seems that the spacetime does not become homogeneous at 
late times, since the coefficient $g^0_{ab}$ of the leading term in the
expansion is not homogeneous. There is, however, a more subtle sense in
which it does become homogeneous. Globally in space there is certainly no 
uniform convergence to a homogeneous metric. This is also the case for
spacetime regions of constant coordinate size in the Gaussian coordinates which
have been used. On a spatial region of fixed physical size, however, things
look different. A region of this kind has a coordinate size which goes
to zero exponentially. Since any metric can be approximated increasingly 
well by a flat metric on a region of ever decreasing size it follows that
on a region of fixed physical size the metric converges uniformly and
exponentially to the de Sitter metric. In this sense the spacetime 
does become homogeneous.

Consider next the flatness problem. If the metric has an asymptotic
expansion of the form given in the last section then  it can be
computed directly that the scalar curvature of the spatial metric
converges to zero exponentially as $t\to\infty$ and this is what we 
want to solve the flatness problem. In fact even more can be said.
The curvature invariants $R_{ab}R^{ab}$ and $R_{abcd}R^{abcd}$
associated with the three-dimensional metric also decay exponentially.
Thus it is not just the scalar curvature which decays; the entire
curvature of the spatial metric decays just as fast. It should be noted
that although the results of \cite{tchapnda03a} and \cite{tchapnda03b} 
give a lot of information on the spacetimes to which they are applicable,
they are apparently not strong enough to give curvature decay.

It is not so easy to address the horizon problem by a simple and
precise mathematical statement. What can be said is the following.
A positive cosmological constant leads to solutions of the Einstein
equations which look like de Sitter space on a long time interval
and a long time interval in de Sitter space does not suffer from
the horizon problem.

\section{Scalar fields} 

As already mentioned in the introduction, an alternative to a positive
cosmological constant as a mechanism for producing solutions of the Einstein
equations with accelerated expansion is a suitable nonlinear scalar field.
Consider a minimally coupled scalar field in a spacetime with vanishing 
cosmological constant. The energy-momentum tensor of the scalar field is of 
the form
\begin{equation}
T_{\alpha\beta}=\nabla_\alpha\phi\nabla_\beta\phi-\left[\frac12
\nabla^\gamma\phi\nabla_\gamma\phi+V(\phi)\right]g_{\alpha\beta}
\end{equation}
where $V$ is a smooth non-negative function, the potential. To see the
connection with a cosmological constant, consider the spatially homogeneous
case. Then the energy density is given by $\rho=T_{00}=\dot\phi^2/2+V(\phi)$
while the pressure is given by $p=T_{11}=\dot\phi^2/2-V(\phi)$. There are 
now different possible regimes. If the kinetic energy is much larger than
the potential energy on a certain time interval then on that interval 
the energy density is approximately equal to the pressure. Thus in a certain
loose sense the matter can be approximated by a stiff fluid, which satisfies 
$p=\rho$. If the kinetic and potential energies are approximately equal
on a certain time interval then the pressure is approximately zero there.
On that interval the matter can be approximated by dust, which satisfies 
$p=0$. Finally, if the potential energy is much larger than the kinetic 
energy then the pressure is approximately equal to minus the energy density.
It is the third case which is related to a cosmological constant. If we 
think of the cosmological constant as a matter field whose energy-momentum
tensor is proportional to the metric then this fictitious matter satisfies
$p=-\rho$. In particular, the pressure is negative and comparable in size
to the energy density and this is what gives rise to accelerated expansion. 

The nature of the dynamics with a nonlinear scalar field depends crucially
on the potential $V$. A useful intuitive picture for guessing
what happens with a given potential is the \lq rolling\rq\ picture. In any
spatially homogeneous spacetime the equation of motion for the scalar 
field is 
\begin{equation}
\ddot\phi-(\tr k)\dot\phi+V'(\phi)=0
\end{equation}
This is similar to the equation of motion of a ball which rolls on the graph
of the function $V$ with variable friction determined by the mean 
curvature $\tr k$. Physical intuition then suggests that the ball should roll
down the slope and settle down in a local minimum of the potential. It
turns out that accelerated expansion eventually stops if the minimum
value of the potential is zero and that for that reason the case of a 
strictly positive minimum is mathematically more tractable.

Depending on how the acceleration of the universe varies with time it may or 
may not be consistent with the simplest model where there is a positive 
cosmological constant and any other matter present satisfies the strong
energy condition and so cannot by itself cause acceleration. If the
obervations are not consistent with acceleration caused only by a 
cosmological constant then the next simplest possibility is the nonlinear
scalar field. Whether a cosmological constant is enough to explain the 
observations does not yet seem to be settled although there is some work
on the problem in the literature. (See e.g. \cite{alam}.)

There have been many suggestions for the form of the potential $V$
in the context of inflation or quintessence but there is no clear winner
at the moment. If there is a scalar field causing cosmological expansion
then we do not know what it is. In these circumstances it makes sense to
study the properties of large classes of potentials. In \cite{rendall04a}
the case of a potential with a strictly positive minimum was discussed.
For spacetimes containing a scalar field of this type and ordinary matter
satisfying the dominant and strong energy conditions it was shown that
there are rather direct generalizations of Wald's theorem \cite{wald} and 
the results of \cite{lee03}.

More specifically, it can be shown under weak assumptions that if the 
potential is bounded below by a positive constant then $V'(\phi)$
tends to zero as $t\to\infty$. Either $\phi$ converges to a finite value 
which is a critical point of $V$ or $\phi$ tends to plus or minus 
infinity. If $\phi$ converges to a finite value and if the corresponding
critical point of $V$ is a non-degenerate local minimum $\phi_1$ then the 
solution has asymptotics like that in Wald's theorem, with $V(\phi_1)$  
playing the role of an effective cosmological constant. The mean
curvature $\tr k$ converges to a constant $-3H_1$.

In \cite{rendall04a} the statement was made that when the potential has a 
non-degenerate positive minimum a solution for which the scalar field 
converges to this minimum has no oscillations. This is misleading and
should be replaced by the statement that the deviation of the scalar 
field from the point where the potential attains its minimum and the 
modulus of $\dot\phi$ decay exponentially as $t\to\infty$. This implies
in particular that $\dot\phi$ is absolutely integrable. The equation for
$u=(\phi,\dot\phi)$ can be written in the form $\dot u=Au+R(t)u$ where
$A$ is a constant matrix and $R(t)$ is a matrix-valued function which decays 
exponentially. If $\beta>9H_1^2/4$ where $\beta=V''(\phi_1)$ then the 
eigenvalues of $A$ are not real. For a generic solution there is an 
oscillation modulating the  leading order exponential decay of the scalar
field.

A natural next step is to look at potentials which are strictly positive
but which are allowed to go to zero at infinity. The best-studied 
case is that of power-law inflation. Analogues of Wald's theorem for
this case were obtained in \cite{kitada93} and extended in \cite{lee04}.
The potential is of the form $V=V_0e^{-\kappa\lambda\phi}$ for a positive 
constant $\lambda$. Here $\kappa$ is a constant which in geometrical
units ($G=c=1$) satisfies $\kappa^2=8\pi$. Accelerated expansion at late 
times is obtained if $\lambda<\sqrt{2}$. If $\lambda$ is greater than 
$\sqrt{2}$ then the expansion is decelerated at late times. In the accelerated 
case the scale factor behaves like a power of $t$ greater than one at late 
times. When $\lambda>\sqrt{2}$ there are exact FLRW models where the scale 
factor is proportional to a power of $t$ which is less than one 
\cite{halliwell}.

For inhomogeneous models there is just one interesting result. In 
\cite{mueller} formal series expansions for spacetimes with power-law 
inflation and matter content given by a scalar field alone were written 
down. It would be interesting to extend the results of \cite{rendall03a} 
for a cosmological constant to this case. The formal expansions are more 
complicated since they can include powers which are any linear combination 
with integer coefficients of one and $\lambda$. This is similar to the
case of a perfect fluid where integer linear combinations of one
and $\gamma$ occur. Note that there is at present no analogue of the
results of \cite{friedrich86} and \cite{friedrich91} known for the
case of power-law inflation. It would also be interesting to extend
the results of \cite{tchapnda03b} to the case of a nonlinear scalar
field. A first step in this direction is a local existence theorem 
for solutions of the Einstein equations coupled to the Vlasov equation
and a linear scalar field which was obtained in \cite{tegankong}.

If the potential is zero somewhere the dynamical behaviour becomes more 
complicated. This is what happens in chaotic inflation. The model case
is that of a massive linear scalar field. There is accelerated expansion
on some finite time interval but it eventually stops, a process known
as reheating. After this the scalar field behaves like dust. At late
times $\dot\phi$ does not decay faster than $t^{-1}$ and so is not 
absolutely integrable. These conclusions are based on heuristic 
arguments \cite{belinskii}.

\section{Relations between perfect fluids and scalar fields}\label{fluid}

A type of matter model frequently used to produce accelerated expansion
is a perfect fluid which violates the strong energy condition. The
equation of state $p=f(\rho)$ satisfies $\rho+3p<0$. In
the simplest case of a linear equation of state $p=(\gamma-1)\rho$
this corresponds to choosing $\gamma<2/3$. Unfortunately $\gamma<1$
means that $dp/d\rho<0$ and so the speed of sound becomes imaginary.
As has been argued in \cite{friedrich00} this suggests that for 
inhomogenous solutions the initial value problem is ill-posed.
The case of homogeneous spacetimes should be thought of as a 
simple and important special case of the problem without symmetry
and if the model makes no mathematical sense without symmetry it is 
suspect.

A solution to this difficulty is the observation that there is a 
certain equivalence between a perfect fluid and a scalar field and that 
the scalar field defines a model which is well-posed without symmetry 
restrictions. Consider first the case of a linear equation of state with 
$0<\gamma<2/3$ and no other matter fields. Suppose that a spatially flat 
FLRW solution is given for a fluid with this equation of state. We look for 
a potential such that the corresponding nonlinear scalar field can reproduce 
the fluid solution. Using the equation of state gives the relation
$\dot\phi^2=\frac{2\gamma}{2-\gamma}V$.
Differentiating this with respect to time gives an equation relating
$\ddot \phi$ and $V'(\phi)$. All terms in this equation have a common factor 
$\dot\phi$. Because $p\ne\rho$ it follows that $\dot\phi\ne 0$ and this 
factor can be cancelled. It follows that
$\ddot\phi=\frac{\gamma}{2-\gamma}V'(\phi)$.   
The Hamiltonian constraint implies that 
$\tr k=-\sqrt{\frac{48\pi V}{(2-\gamma)^2}}$. 
Putting all this information
into the equation of motion for the scalar field gives the equation
$V'=-\sqrt{24\pi\gamma}V$. Solving this equation shows that 
$V=V_0 e^{-\sqrt{24\pi\gamma}\phi}$. Thus the only kind of potential
which can work is the one we have already seen for power-law inflation,
with $\lambda=\sqrt{3\gamma}$. The range of values of $\lambda$ which
occurs is exactly that which we already saw. It can be shown that 
this potential really does reproduce the fluid solution. To see this,
notice that the initial data which must be chosen for the scalar field
are uniquely determined by the data for the fluid. The quantities $p$
and $\rho$ defined from this scalar field satisfy the Euler equations
since the energy-momentum tensor of the scalar field is divergence-free.
Hence they agree with the fluid density and pressure everywhere. Note
that this procedure does not extend to models which are homogeneous 
but not isotropic.

The above analysis can be generalized to other equations of state.
Consider again the case of a perfect fluid and no other matter fields.
Some general assumptions will be made on the equation of state to
make a smooth and complete discussion possible. It should, however,
be noted that the considerations which follow may be usefully applied
in more general situations. Here it is assumed that $dp/d\rho<C_1<1$ for a
constant $C_1$ and $p/\rho>C_2>-1$ for a constant $C_2$. Note that for
any nonlinear scalar field $|p/\rho|\le 1$. For a general equation of 
state the relation
\begin{equation}\label{implicit}
\frac12\dot\phi^2-V(\phi)=f\left(\frac12\dot\phi^2+V(\phi)\right) 
\end{equation}
must be analysed. This can be rewritten in the form 
$F(\frac12\dot\phi^2,V(\phi))=0$. Suppose that we have one solution of this 
equation. The implicit function theorem gives the existence of a 
function $g$ which satisfies $F(g(V), V)=0$ for $V$ close to 
its value in the original solution. This is because the partial derivative 
of $F$ with respect to its first argument is non-zero. The function $g$
satisfies the relation
\begin{equation}
g'(V)=\frac{1+f'(g(V)+V)}{1-f'(g(V)+V)}
\end{equation}
As a consequence the derivative of the locally defined function $g$ remains
bounded on its domain of definition and $g$ can be extended to a longer
interval provided it does not tend to zero at the endpoint of the interval.
If $g$ tended to zero then this would imply that $p/\rho\to -1$, in
contradiction to what has been assumed concerning the equation of state.
It follows that the relation (\ref{implicit}) can be inverted globally
to give $\dot\phi^2=2g(V)$. Following the same steps as in the case of
a linear equation of state gives the equation
\begin{equation}
V'(\phi)=-\frac1{1+g'(V)}\sqrt{48\pi g(V)(g(V)+V)}
\end{equation}

An exotic fluid model for accelerated cosmological expansion is the 
Chaplygin gas \cite{kamenshchik} with equation of state $p=-A/\rho$
for a positive constant $A$. It satisfies $dp/d\rho>0$ but violates
the dominant energy condition for $\rho<A$, since in that case $p/\rho<-1$.
It is ruled out by the assumptions made above. It turns out, however 
that there are cosmological models with this equation of state where 
$\rho>A$ everywhere so that this difficulty is avoided. The calculations
as above can be done for the Chaplygin gas assuming the inequality
$\rho>A$. The result is surprisingly simple. The potential is
given by $V(\phi)=\frac12\sqrt{A}(\cosh \sqrt{24\pi}\phi
+\frac1{\cosh \sqrt{24\pi}\phi})$.
Thus a potential is obtained which has a strictly positive lower bound
and it satisfies the hypotheses of the theorems of \cite{rendall04a}. 
Thus for the scalar field corresponding to the Chaplygin gas detailed
information is available about late time asymptotics. Unfortunately,
because of the fact that the transformation to the scalar field picture
is not globally defined, it is not possible to immediately deduce
full information on the late-time dynamics for the Chaplygin gas.
It should also be remembered that the correspondence with a scalar
field does not apply to solutions of the Einstein equations with a 
Chaplygin gas which are homogeneous but not isotropic.

Sometimes it is desirable to parametrize the degrees of freedom in a 
cosmological model with fluid in a way which is different from that using
the equation of state. An important example of this is the machinery 
required to compare supernova observations with theoretical models.
This will now be sketched. If we know both the apparent and intrinsic 
brightness of a source then we can compute its distance. (Technically,
what can be computed is the so-called luminosity distance.) Consider
a spatially flat FLRW model. Then the redshift $z$ of an object is
given in terms of the scale factor by $1+z=a(t_o)/a(t_e)$, where $t_o$
is the time at which the light from the object is observed and $t_e$ 
the time at which it is emitted. In a model of this kind the luminosity 
distance can be computed to be $D_L=ra(t_o)(1+z)$, where $r$ is the 
spatial separation between the wordlines of observer and emitter 
as measured in standard coordinates. Let $H=-\tr k/3$, the Hubble
parameter. If the luminosity distance and Hubble parameter are expressed
in terms of redshift then the following relation results \cite{statefinder}:
\begin{equation}\label{hubble}
H(z)=\left[\frac{d}{dz}\frac{D_L(z)}{1+z}\right]^{-1}
\end{equation}
Supernova data provides points on the curve $D_L(z)$ and the equation 
(\ref{hubble}) in principle then determines $H(z)$.

Let us ignore complications due to having only discrete data and suppose
we know the function $D_L(z)$ exactly. It will now be shown how the scale
factor $a(t)$ can be reconstructed.  Firstly, $H(z)$ can be computed 
using (\ref{hubble}). An elementary computation shows that
$dt/dz=-H(z)(1+z)^{-1}$. Integrating this gives $t$ as a function of $z$ 
and inverting this gives $z$ as a function of $t$. Thus $H(t)$ can be 
determined. Integrating once more gives $a(t)$. In practise, in order 
to distinguish between different theoretical models, an ansatz is made 
for $H(z)$ containing some parameters and a best fit analysis of the data
is carried out to obtain values for these parameters.

\section{Tachyons and phantom fields}

The ordinary scalar field we have considered up to now can be derived from 
a Lagrangian with density $\nabla_\alpha\phi\nabla^\alpha\phi+V(\phi)$.
Recently dark energy models have been considered where the Lagrangian 
density is a more general nonlinear function 
$p(\nabla_\alpha\phi\nabla^\alpha\phi,\phi)$. This is known as 
$k$-essence \cite{armendariz}. A great advantage of the ordinary
nonlinear scalar field is that it is guaranteed to have well-behaved
dynamics in the full inhomogeneous case. The Cauchy problem is 
always well-posed. (This is even true if the potential is allowed
to be negative.) In contrast, $k$-essence models need not have 
a well-posed local Cauchy problem. The equation of motion of the
scalar field need not be hyperbolic. An additional complication is
that since the equations are in general quasilinear rather than
semilinear, the scalar field may develop shocks. In this case there
is an additional source of singularities supplementing the familiar
ones in general relativity. A useful discussion of the some of these
points, and the question of which energy conditions are satisfied by 
$k$-essence models, can be found in \cite{gibbons}. The models which
violate the dominant energy condition are called phantom or ghost models.

An interesting example is given by the case where the function $p$
is given by $V(\phi)\sqrt{1+\nabla_\alpha\phi\nabla^\alpha\phi}$, which
is known as the tachyon or tachyon condensate. Note that although the word 
\lq tachyon\rq\ originally denoted a particle which travels faster than 
light, the tachyon field considered here has no superluminal propagation. 
All characteristics of the equation lie inside the light cone. The tachyon 
condensate corresponds to an effective field theory for a large collection
of tachyons. Consider now the special case where $V(\phi)$ is identically 
one. Then provided the gradient of $\phi$ is timelike this model is 
equivalent to a special case of the Chaplygin gas. To see this it suffices 
to define the four-velocity of the fluid by
\begin{equation}
u^\mu=\frac{\nabla^\mu\phi}{\sqrt{-\nabla_\alpha\phi\nabla^\alpha\phi}}.
\end{equation}
This velocity field is irrotational. The equation for a Chaplygin gas in
four-dimensional Minkowski space also describes a timelike hypersurface of 
zero mean curvature (a membrane) in five-dimensional Minkowski space. 
Questions of global existence for these equations have been studied in 
\cite{lindblad}.

\section{Closing remarks}

This paper gives a general introduction to the subject of 
cosmological models with accelerated expansion, taking a mathematical
point of view. After some basic concepts have been introduced, the 
relevant physical background on inflation and quintessence is outlined.
After this, various existing mathematical results in the case of a
positive cosmological constant are presented. They are then confronted
with the physical motivation. The exposition continues with a review
of results in the case where the cosmological constant is replaced by
a nonlinear scalar field. Some interesting open problems are mentioned.
There are close relations between models with scalar fields and models
with perfect fluids whose equation of state is more or less exotic.
Some of these connections are explained. Following this it is explained
how scalar fields defined by Lagrangians which are non-linear in the
first derivatives give rise to models (known as $k$-essence) which
various connections to both more conventional scalar fields (which 
are linear in derivatives) and perfect fluids.

At this moment new observations on cosmic acceleration are stimulating a 
vigorous model-building activity. One aspect of this is that if string
theory is a theory of everything then it should, in particular, be
able to explain dark energy. It is thus natural that string theory 
should be one of the main sources of new models. There are many
models which are not touched on at all in this paper, in particular
those coming from brane-world scenarios \cite{maartens} or loop
quantum cosmology \cite{bojowald}. We have taken a conservative
strategy which covers some of the models which are easier to
understand mathematically. Even with these limitations we could only 
treat a few aspects of the subject. A useful task for mathematical
relativity is to establish clear definitions of the various models
and to identify interesting dynamical issues concerning the solutions.
Another task is to systematize the web of relations which exists 
relating different models and to determine which of them are (in an
appropriate sense) really different. Apart from its pedagogical aspects
this paper is intended to be a step towards meeting these challenges.

\section{Acknowledgements} 

I thank Yann Brenier for useful discussions. I gratefully acknowledge the 
support of the Erwin Schr\"odinger Institute, Vienna, where part of the 
research for this paper was carried out.

\end{document}